\documentclass[preprint,showpacs,preprintnumbers,amsmath,amssymb,aps]{revtex4}

\usepackage{graphicx}

\newcommand{\beq}{\begin{equation}}
\newcommand{\eeq}{\end{equation}}
\newcommand{\kpc}{\, {\rm kpc} }
\newcommand{\etal}{{\it et al.}}
\newcommand{\MNRAS}{{\it Mon. Not. R. Astron. Soc.} }
\newcommand{\ApJ}{{\it Astrophys. J.} }
\newcommand{\Nature}{{\it Nature} }

\begin{document}

\title{Mapping the dark trajectory of the Bullet Cluster }

\author{X. J. Bi}
\affiliation{Key laboratory of particle astrophysics, IHEP,
Chinese Academy of Sciences, Beijing 100049, P. R. China}

\author{H.S. Zhao}
\affiliation{SUPA, School of Physics and Astronomy, University of
St Andrews, KY16 9SS, Fife, UK}

\date{\today}

\begin{abstract}
We model the electrons/positrons produced by
dark matter annihilations in the colliding
galaxy cluster system 1E0657-56 (the bullet cluster).
These charged particles, confined by the Magnetic
field, clearly trace the path of the bullet, which passes through the
main cluster with a speed of 3000-5000 km/s.  Adding the effect of dark matter substructure (subhalos) in each cluster we find the annihilation rate is enhanced
greatly and the density of positrons in the trail is similar to that
within the bullet cluster.  This opens the door to
unambiguous detection of the annihilation signal through the SZ effect, at significant separation from the thermal electrons.

\end{abstract}
\maketitle


The nature of dark matter particles is one of the most outstanding puzzles
in particle physics and cosmology. A massive effort has been made in the last decade to unravel the mystery, but to little avail. Numerous theoretical models have been proposed, among which
the most attractive scenario involves weakly interacting
massive particles (WIMPs). A viable way to identify
WIMPs is to search their annihilation products, such as $\gamma$-rays,
neutrinos, positrons or anti-protons. Up to now most efforts
have focused on looking for the signals within the Milky Way.
We know there are large quantities of dark matter residing in galaxy clusters.
However, they are typically too distant to observe these annihilation signals
directly.
One viable way to look for the effects of dark matter annihilations
in clusters is to compute the SZ effect produced by the non-thermal
electrons/positrons which scatter
off the background CMB photons \cite{cola2006}, but are not
weakened by the long distance.

However, the SZ effect from the thermal electrons in the cluster X-ray gas is usually much greater than the SZ effect from the
non-thermal electrons produced from dark matter annihilations
and leads to difficulties disentangling them. Colliding galaxy clusters provide one of the best systems in nature
where we see a clear offset of baryonic material concentrations
and the gravitational centres, which are generally interpreted
as direct evidence of the long-speculated dark matter.
For example, the bullet cluster system 1E0657-56 (\cite{Clowe}, \cite{Angus1}) is such a
system with clear spatial separation of the X-ray gas and the
dark matter halos, which offers an ideal place to search for
signals of dark matter annihilation.

For the bullet cluster, we observe, at optical wavelengths, two concentrations of galaxies
with the smaller one passing through the larger one.
Clearly the two systems are merging with one another. X-ray observations show
a hot gas envelop joining both clusters, and an intense bow-shocked
structure in between the optical centres.  Looking past these
clusters, we observe systematic
weak distortions of the shapes of the background galaxies,
with the signal being concentrated on the galaxies and not the X-ray gas, which is impossible if there is not more dark matter co-existing with the galaxies in some
collisionless form.


The bow-shock observed in the gas distribution of the sub cluster, implies a supersonic motion of the shock of about 5000km/s
in the $X$-direction perpendicular to the line of sight.
The underlying dark matter might move somewhat slower, about 3000km/s
in the recent simulation \cite{Springel}, but see \cite{Angus2} for a discussion.  The centre of
the X-ray emitting gas is about 300 kpc to the left (East) of the bullet,
and 400kpc to the West of the main cluster.

A halo of synchrotron emission has been observed at radio wavelengths
for this cluster \cite{Liang}, which might be associated
with the elongated emission of the X-ray gas. However, there is no indication
for any enhancement near the DM clump locations.
A SZ map of 1E0657-56 have also been obtained with ACBAR \cite{Gomez},
which again does not show any enhancement,
at the centres of the dark halos,
although the resolution $\sim 4.5' \sim 100\kpc$ at the
cluster's redshift ($z=0.3$).
Recently Colafrancesco et al. \cite{cola2007} made a new calculation
of the SZ effect by the nonthermal electrons from dark matter
annihilation. They conclude that a detection of such effect
is feasible in the future observations.

In the present work we propose a new effect in this system
which can clearly show their origin from dark matter annihilation.
Since the dark halo of the bullet cluster is offset from the thermal
gas and moving with a huge speed within the main cluster
we expect the electrons/positrons from dark matter annihilation in
the bullet cluster
will be confined by the magnetic field, which is frozen with the gas,
and spread over a long trail.
Further we take the subhalos into account when calculating the
dark matter annihilation, which can greatly enhance the annihilation signal.
The effect of subhalos also makes the electron density in the trail
almost the same magnitude as that within the bullet cluster.
Therefore we expect the SZ effect from this trail to be of the same order
as from the main and bullet clusters and detectable in the future
observation.



A promising WIMP candidate is the neutralino, the lightest supersymmetric
particle, which is neutral and generally stable. Neutralinos can
annihilate into a pair of quarks, leptons or gauge bosons, which
decay to electrons/positrons and some other states finally.
In the present work we assume neutralinos account for all dark matter in the universe.

The source function of electrons and positrons from DM annihilations
can be written as
\begin{equation}
Q(E_{e},{\bf r})=\frac{\langle\sigma v\rangle}{2m_{\chi}^2}\frac{{\rm d}N}{{\rm d}E}
\rho^2({\bf r}) = f(E_e) \rho^2({\bf r}),
\label{positronsource}
\end{equation}
where $\sigma$ is the electron/positron generating cross-section,
${{\rm d}N}/{{\rm d}E}$ is the electron/positron spectrum produced by one annihilation
from a pair of neutralinos and
$\rho({\bf r})$ is the neutralino density distribution in space.

The factor $f(E_e)$ in the source term is calculated in
the minimal supersymmetric standard
model (MSSM) by performing a random scan in the SUSY parameter space
using the software package DarkSUSY \cite{Gondolo00}.
Under some simplification assumptions only seven most relevant
SUSY parameters are kept, as done in DarkSUSY \cite{Gondolo00}. 
We require the models to satisfy the constraints from colliders and cosmology.
In particular, the relic density
$0 < \Omega_\chi h^2< 0.118 $ is required,
which is lower than the 3$\sigma$ upper bound from
the cosmological observations \cite{Spergel03}.
For the small relic density we assume a nonthermal production
mechanism \cite{zhang}. Among the randomly produced 1000 models
we choose a model which has a large branching ratio to
electrons and positrons. 
Here we produce only $1000$ models in
order not to introduce too large fine tuning of the SUSY parameters. 

The second factor $\rho^2({\bf r})$ in Eq. (\ref{positronsource})
is determined by the dark matter distribution.
We model the two clusters with NFW profiles, as shown in \cite{Springel} and adopt the following structure parameters:
$M_{vir}= 1.5\cdot 10^{15} M_{\odot}$, $R_{vir} = 2.3$ Mpc
and $c_{vir}= R_{vir}/r_s = 3$ for the larger (East) DM halo;
$M_{vir} = 1.5 \cdot 10^{14}
M_{\odot}$, $R_{vir} = 1.070$ Mpc and $c_{vir}= 7.2$ for the
``bullet'' (West) DM clump.  This mass profile approximates the signal
of weak gravitational lensing satisfactorily.

\begin{figure}
\resizebox{7.5cm}{!}{\includegraphics{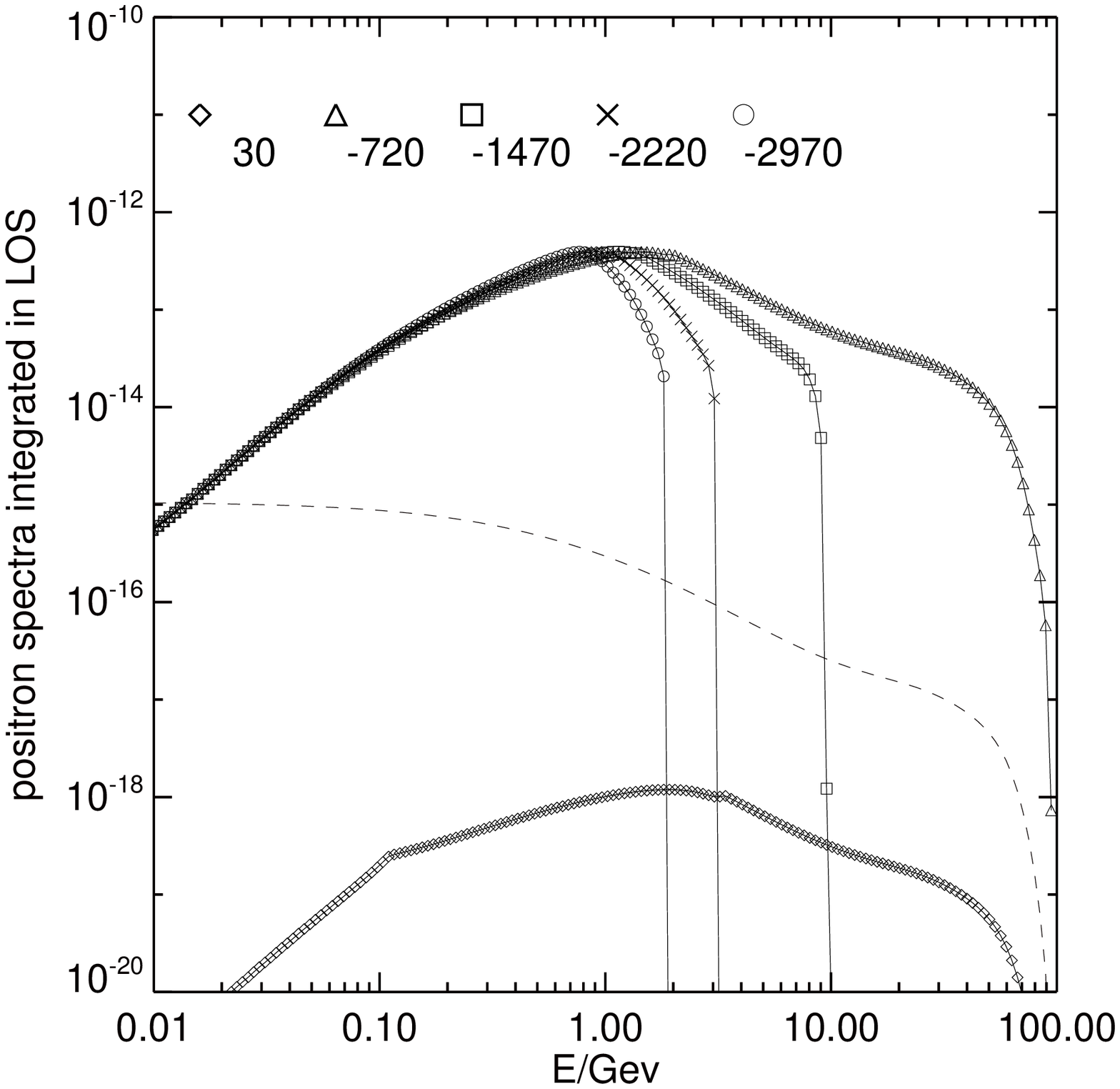}}
\resizebox{7.5cm}{!}{\includegraphics{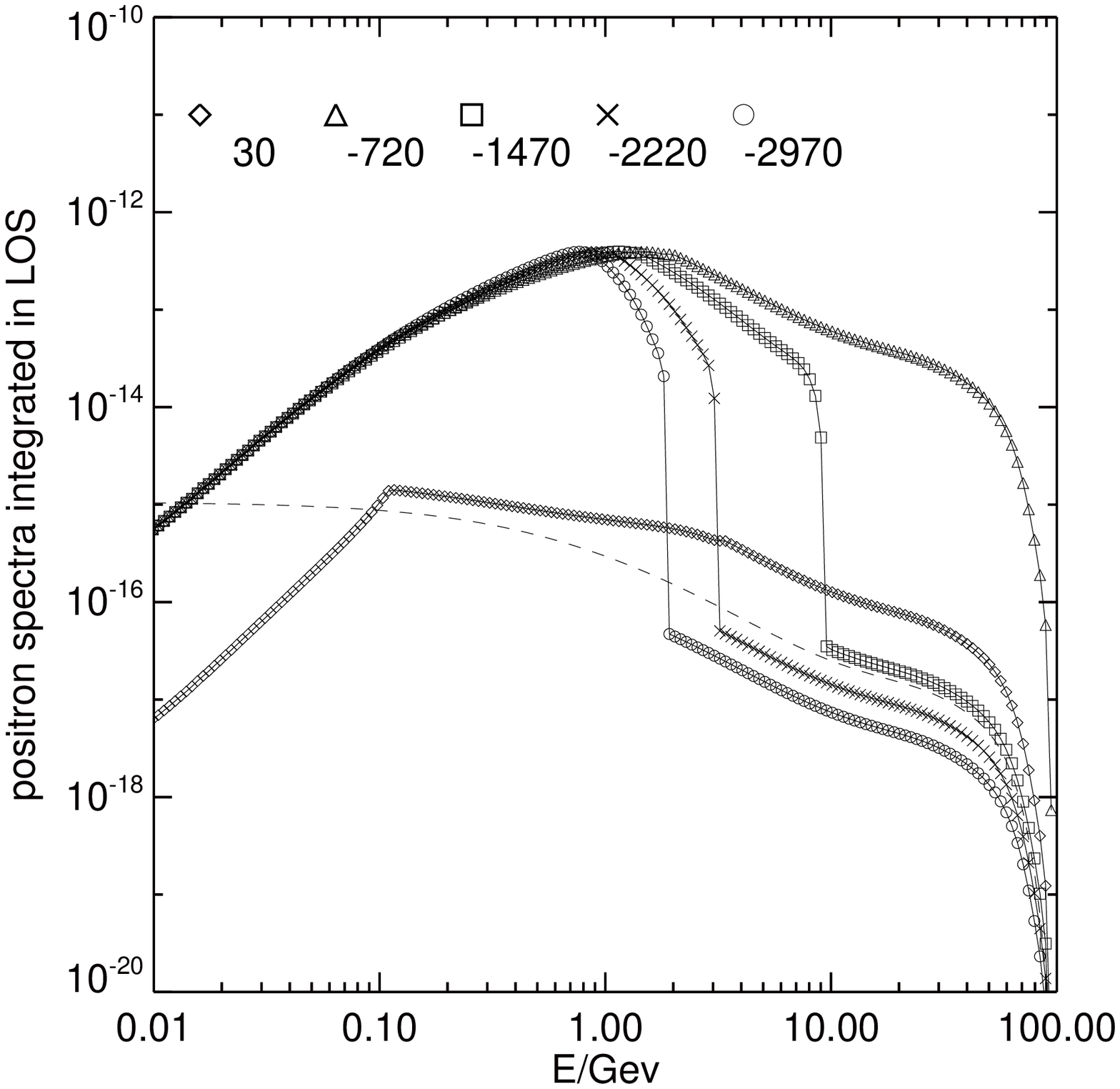}}
\caption{\label{spec}
Shows the line-of-sight integrated
spectrum $\int_{-\infty}^{\infty} E^2 n_e(E,x,y=0,z) dz$
for different positions on the X-axis ($x=30,-720,-1470,-2220,-2970$kpc).
Here $x$, $y$, $z$ are the directions of the bullet's motion, distance
from the cluster centers and along the ling of sight.
Here we compare the steady-state spectrum (dashed line which levels off at low energy, $x=30$kpc) with
the moving model (left) and moving model with substructure (right).
The spectrum is softer near the ``tail'' of the bullet.
}
\end{figure}

For substructures within each cluster we have the number density of
substructures with mass $m_{\mbox{sub}}$ at the position
$r$ to the cluster center as \cite{Diemand04}
\begin{equation}
\label{prob}
n(m_{\mbox{sub}},r)=n_0 \left(\frac{m_{\mbox{sub}}}
{M_{\mbox{vir}}}\right)^{-1.9} (1+(r/r_H)^2)^{-1}\ ,
\end{equation}
where $M_{\mbox{vir}}$ is the virial mass of the cluster,
$n_0$ is the normalization factor determined by requiring that
about $50\%$ of the dark matter is enclosed in the subhalos \cite{diemand2005}.
We take the minimal subhalo to be as light as $10^{-6} M_\odot$ as shown in
the simulation by Diemand et al. \cite{diemand2005} and the
maximal subhalo about $0.01 M_{vir}$.
To determine the profile of each subhalo we adopt the semi-analytic model
of Bullock et al. \cite{Bullock01} to give the
concentration parameter $c_{\mbox{vir}}$ as a function of the subhalo
viral mass. Simulation also shows \cite{Bullock01} that subhalos in dense
environment usually have larger concentration parameters than the corresponding
distinct halos with same mass. Here we take a factor of $~ 2$ larger than the
concentration parameter given by the Bullock model.

Taking the subhalos into account, the density squared term in Eq.
(\ref{positronsource}) is replaced by
\begin{equation}
\rho^2({\bf r}) \to \langle \rho^2({\bf r}) \rangle
= \rho^2_{\mbox {smooth}}({\bf r}) + \langle \rho^2_{\mbox{sub}}
({\bf r}) \rangle,
\end{equation}
where $\langle \rho^2_{\mbox{sub}}({\bf r}) \rangle$ means the average
density square of subhalos according to the subhalo distribution probability.
Since there is no correlation between the mass and spatial distribution
in Eq. (\ref{prob}) we get $\langle \rho^2_{\mbox{sub}}({\bf r}) \rangle$
at the position ${\bf r}$ from an integral over mass
\begin{eqnarray}
\langle \rho^2_{\mbox{sub}}({\bf r}) \rangle &=&
\int_{m_{{min}}}^{m_{{max}}} n(m_{\mbox{sub}},r)
\left( \int \rho_{\mbox{sub}}^2 {\rm d}V_{\mbox{sub}} \right)
\cdot {\rm d}m_{\mbox{sub}} \nonumber\ \ ,
\label{averrhosq}
\end{eqnarray}
where $\rho_{\mbox{sub}}$ refers to the density of the subhalo at ${\bf r}$ and
$V_{\mbox{sub}}$ is its volume.


When the bullet crosses the main cluster, the dark matter density at each
point is the sum of the two individual clusters and changes with time.
Therefore the  annihilation signal
is enhanced due to the superposition of the dark matter density of
the two clusters.
A detailed match of CDM models with the Mach-cone in the X-ray suggests
that the two CDM halos are presently moving with about 3000km/s to each other.
The speed was initially 1800km/s at the contact of the two halos and
was maximized to 4000km/s when the two cluster centers coincide.
Overall we take an average of 3000km/s as suggested
in the recent model of Springel and Farrar \cite{Springel}.


The electrons/positrons from dark matter annihilation will undergo
the processes of diffusion and energy loss in the intracluster medium.
The diffusion equation is
\beq
\frac{\partial}{\partial t} \frac{dn_e}{dE} =  \frac{\partial}{\partial E}
\left( b(E)  \frac{dn_e}{dE} \right) + Q(E,t;{\bf r})\ \ ,
\label{trans}
\eeq
where we have dropped the spatial diffusion term 
since the diffusion time scale is longer than that of
energy loss for electrons/positrons with $E \gtrsim 10 MeV$ \cite{cola2006}, 
and $b(E)= - \frac{dE}{d\tau}$ is the cooling
function. It should be noted that the source function here is time-dependent
with the motion of the bullet cluster, which is
different to that in Eq. (\ref{positronsource}).

The energy loss of electrons/positrons comes mainly from the inverse 
Compton effect and synchrotron radiation \cite{cola2006}.
Neglecting the Coulomb scattering and bremsstrahlung, which are not important
for the energy range that we are interested in, we get \cite{cola2006}
\beq
b(E) =  b_0 \left( \frac{E}{1 GeV} \right)^2 =\left( b_{IC}^0 + b_{syn}^0 B_\mu^2 \right)
 \left( \frac{E}{1 GeV}\right)^2 \ ,
\eeq
with a Magnetic field $B_\mu\approx 1\mu G$ inside the virial radius
of the main cluster.
The simple form of $b(E)$ leads to a simple analytical solution of the transport
equation (\ref{trans}) given bellow.
(The full solution of the diffusion equation with time dependent source terms
is given by Baltz and Wai \cite{baltz}.)

The source term $Q$ can now be expressed as
\beq
Q(E,t; {\bf r}) = \rho_X^2(t;  {\bf r}) f(E)\ ,
\eeq
with the underlying dark matter density given as
\begin{eqnarray}
\rho_X(t; {\bf r}) &=& \sum_{i=1}^{2} \rho_{NFW}(|{\bf r}-{\bf r}_i(t)|)
                + \rho_{sub}(|{\bf r}-{\bf r}_i(t)| \\
 &=& 0, \qquad t \le 0
\end{eqnarray}
which is decomposed into a smooth term and a substructure term, and summed
over both clusters $i=1, 2$.  The coordinates of the cluster centers in the past
are given by
\beq
{\bf r}_i(t) = {\bf r}_{i,now} + \int_{t}^{t_{now}} {\bf V}_i dt',
\eeq
where $ {\bf r}_{i,now}$
are the present positions of the cluster centers, and ${\bf V}_i$
are their velocities in the past.  We take the simple assumption of a
head-on collision with a constant speed $V=3000$km/s for the bullet
and zero for the main cluster.

The solution at any epoch $t$, position ${\bf r}$ and energy $E$
is
\begin{equation}
n_e(E, {\bf r}, t) = 
\frac{1}{b(E)} \int_E^{m_\chi} dE_e  Q(E_e,t_e; {\bf r})
\end{equation}
with the emitting time
\begin{equation}
t_e = t - \tau(E \leftarrow E_e) = t - \int_{E}^{E_e} \frac{dE'}{b(E')}
=t+\frac{1}{b_0}\left(\frac{1}{E_e}-\frac{1}{E}\right)\ ,
\end{equation}
where $\tau$ is the time to cool an electron from $E_e$ to $E$.
The initial condition is taken as $n_e = 0$ at about 10Gyrs ago.

A general practice has been to assume the electrons reach equilibrium and have
a static solution of Eq. (\ref{trans})
with $\frac{\partial}{\partial t} n_e(E, {\bf r}, t) =0$.
However, we will show below that the time dependent spectrum is different
from the static solution adopted in previous works \cite{cola2007}.
This is because the time scale of energy loss for low energy electrons 
is longer than the
source production time and has not reached its static state, as shown
in \cite{cola2006}.



\begin{figure}
\resizebox{7.3cm}{!}{\includegraphics{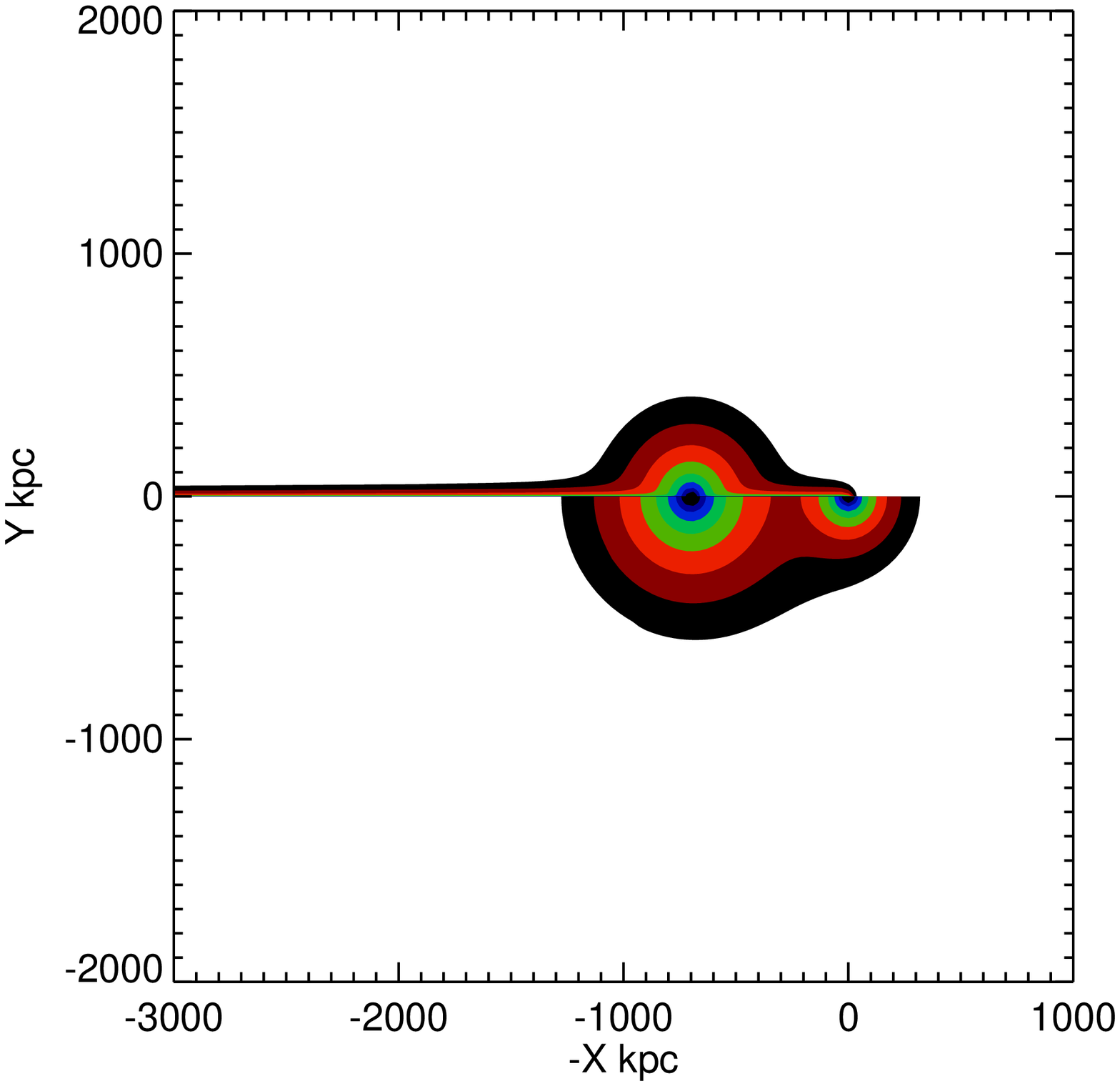}}
\resizebox{7.3cm}{!}{\includegraphics{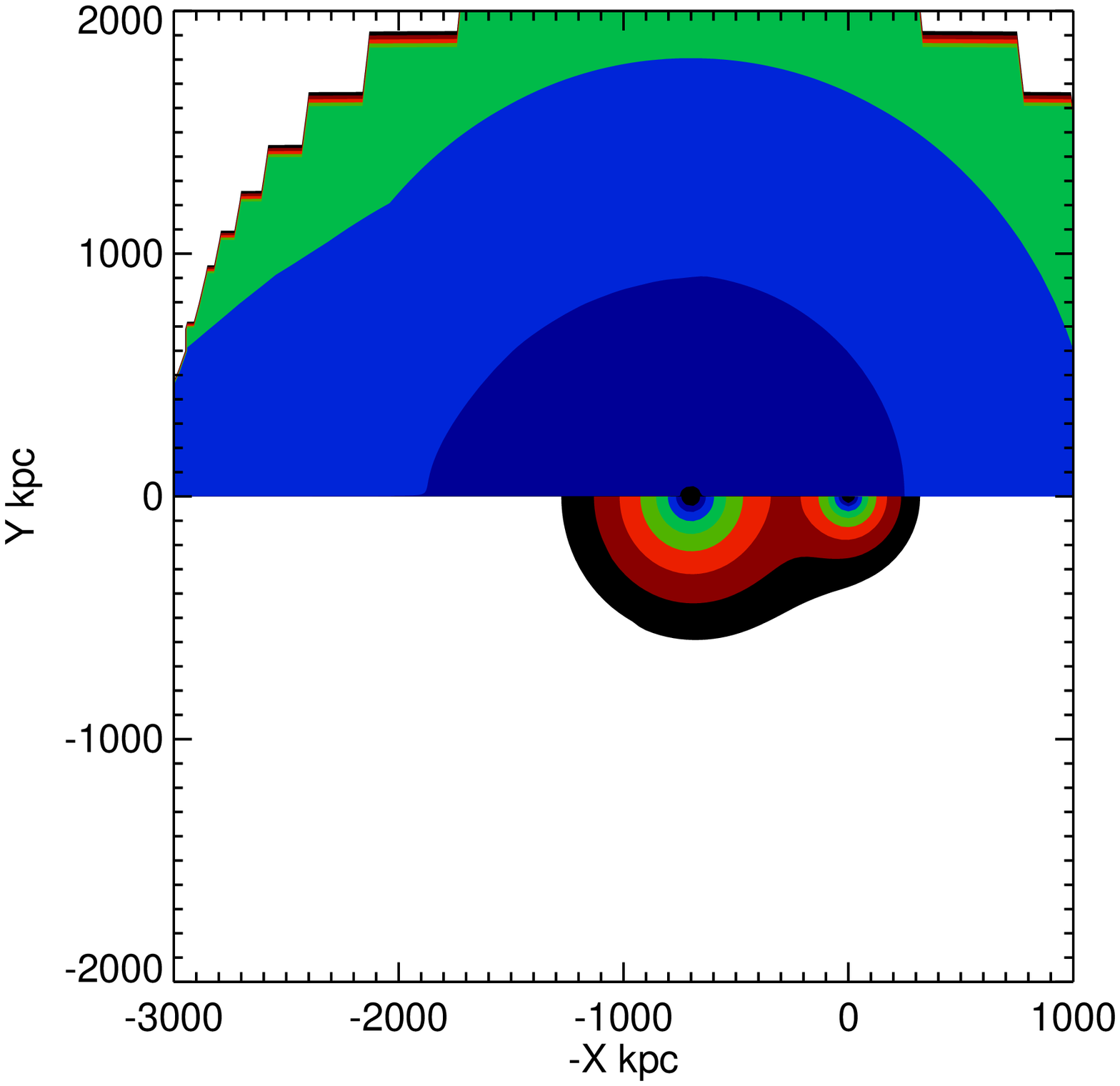}}
\caption{\label{compare}
shows the $e^{-}e^{+}$ pressure distribution $p(x,R)=\int n_e(E,x,R) E dE$
in the XR plane (no projection, $R=\sqrt{y^2+z^2}$). Here we compare
the standard double-NFW steady-state model
(lower half of two panels) with our moving bullet model (upper left half with ``tail'') and
moving bullet with substructure model (upper right half).
The bullet is the smaller cluster at $X=Y=Z=0$, is moving further
to the right with 3000km/s. Contour spacing is 0.5 dex.
}
\end{figure}

In Fig. \ref{spec} we show the electrons/positrons spectra at different
positions compared with the static spectrum. At high energy our spectra
is the same as that of the static solution at the corresponding positions
while at low energies our spectra 
coincides with that of the original spectrum multiplied by the time the source
has existed. This means for low energy electrons the energy loss is negligible.
We noticed an interesting feature of the spectra at the tail which
have a break energy, above which the electrons
lose energy rapidly and reached equilibrium, while below which the electrons
lose energy slowly and leave the trace of the bullet passing with much higher
fluxes. The effect of substructures enhances the annihilation signal 
greatly at large radii. However tidal force disrupts subhalos near the cluster
certer 
\cite{yuan}. Therefore the signal at $X=30$ kpc is enhanced greatly. At other
positions with high energies signals are also greatly enhanced since the static
spectra at these positions are enhanced. At low energies the spectra are
not determined by the local density, but by the accumulating history and
dominated by the flux of the cluster center when it passes these positions.
(Note we are considering the positions with $Y=0$).
Therefore substructures do not change the low energy spectra. (Certainly low
energy spectra are enhanced at positions with $Y\neq 0$, see Fig. \ref{compare}.)

A good indicator of the SZ effect is the non-thermal pressure
\beq
P({\bf r},t) = \frac{1}{3} \int_0^{E_{max}} n(E,{\bf r},t) E dE
\eeq
for the relativistic electrons.
The trail caused by the bullet's motion shows up in projection. 
The inclusion of the subhalos significantly enhances the
annihilation and makes the trail more obvious.
In Fig. \ref{compare} we show contours of the pressure in
different cases: two clusters with the static spectrum solution
of Eq. (\ref{trans}) as adopted in \cite{cola2007} without motion, 
two clusters with motion
with and without contribution from subhalos.
From the figures we clearly see the trail produced by the motion 
of the bullet cluster.
We also notice that the strength of pressure at the trail 
decreases further from the
bullet cluster. However, considering the contribution from the 
subhalos reduces the attenuation.



The non-thermal relativistic positrons/electrons from annihilation
distort the CMB spectra at frequencies much higher than the usual
SZ-effect due to keV thermal electrons.  
We note that the original conclusion of Colafrancesco
et al. \cite{cola2007} that the non-thermal SZ signals of the DM
are both separable from the thermal SZ-effects and detectable with future
experiments remains valid for our model: the earlier model likely
underestimated the boosting factor of annihilation
due to the internal substructures
inside each NFW cluster halo.  More importantly angular resolution is
less critical because of the larger separation (about 2000 kpc, or 13 arcmin)
between the positron tail from the thermal electrons; in
Colafranscesco et al's case, the separation is only about 200 kpc, which is
near the resolution of ground experiments like
South Pole Telescope \cite{ruhl},
Actama Cosmology Telescope \cite{Kosowsky}, and balloon-borne
experiments like OLIMPO \cite{Masi}.  The
relaxed thresholds for sensitivity and angular resolution should
increase the chance of annihilation-induced SZ effects being detected.
The detection of a trail-like SZ signal would compound the proof of
dark matter in the bullet cluster \cite{Clowe}.

In conclusion we consider a new effect to search the WIMP annihilation
signal in the colliding bullet cluster system 1E0657-56.  We find the motion of the bullet cluster will leave a clear
trail of the annihilated electrons/positrons, which may lead to
an observable SZ effect.  In particular, taking the contribution from
subhalos into account, considerably boosts the annihilation rate at
large radii, meaning the pressure of the non-thermal electrons
deceases slowly after the bullet has passed.

\begin{acknowledgments}

H.S.Zhao thanks PPARC and CAS, and X.J. Bi thanks Royal Society.
X.J. Bi is supported by the NSF of China under the grant Nos.
10575111, 10773011 and supported in part by the Chinese Academy of Sciences
under the grant No. KJCX3-SYW-N2.

\end{acknowledgments}

\end{document}